# Nanotube-based systems for broadband optical limiting : towards an operational system

N. Izard[1,2], D. Riehl[1], E. Anglaret[2]
1-Centre Technique d'Arcueil, DGA, Arcueil, France
2-Groupe de Dynamique des Phases Condensées, UMR CNRS 5581,
Université Montpellier 2, Montpellier, France

**ABSTRACT**
Nanotube-based systems are good candidates for optical limiting against broadband laser pulses. We explore new routes to improve their limiting performances. We show that the diameter of the nanotubes is a key factor to control the performances. On the other hand, we demonstrate that chemically modified nanotubes can be mixed with organic chromophores, leading to high performance composite limiting systems which are particularly efficient in the nanosecond regime due to the cumulative effects of nonlinear scattering and multiphoton absorption.

**INTRODUCTION**
Carbon nanotubes are fascinating molecular systems which display original electronic and optical properties. When homogeneously dispersed in organic solvents, nanotubes display optical limiting properties [1]. The main limiting phenomenon is a strong nonlinear scattering due to heating of the nanotubes and the concomitant growth of solvent bubbles at low fluences and carbon vapor bubbles at larger fluences [1]. However, the performances of nanotubes still don't reach those required for effective, auto-activated, protection of eyes against laser beams.
In this paper, we explore two routes for optimizing the optical limiting performances of nanotube-based systems. First, we study the relation between nanotube structure and limiting performances. Second, we develop composite systems by mixing nanotubes with organic chromophores, and we show that the association of nonlinear scattering from nanotubes and two-photon absorption from chromophores is an effective way to prepare high performance optical limiting systems.

**EXPERIMENTAL**

We studied both single wall (SWNT) and multiwall carbon nanotubes (MWNT). SWNT were prepared by the electric arc technique (from Nanoledge, Inc). MWNT were prepared by CVD techniques using different physico-chemical parameters in order to control both the length and diameter of the nanotubes (from Nanolab, Inc). For preparing composites, high performance two-photon absorbers (TPA) were synthesized in the group of Chantal Andraud at Ecole Normale Supérieure de Lyon. The results we present below are for a fluorene pentamer (FMP).
Aqueous suspensions of SWNT and MWNT were prepared using SDS, an ionic surfactant. When the suspensions are sonicated at low powers for a few hours, homogeneous and stable suspensions are obtained. In the case of SWNT, the nanotubes remain aggregated into bundles, as stated by electron microscopy and Raman measurements (not shown). By contrast, high power sonication (500 W during 15 minutes) leads to exfoliation of the bundles [2]. After an ultracentrifugation step (120000 g during 4 hours), aqueous suspensions of individual SWNT are formed, as stated by electron microscopy, photoluminescence [3] and photon correlation spectroscopy [4] techniques. Suspensions of grafted nanotubes were also prepared in chloroform (see details below). The structure of the nanotubes, in powders and in suspensions were extensively studied by various spectroscopic techniques [3,5].

For optical limiting experiments, the concentration of nanotubes in the suspensions was adjusted in order to get cells with 70% transmittance at low fluences, which corresponds to nanotube concentrations of about 100 ppm. Optical limiting measurements were performed using two differents lasers sources and test beds : (i) a Q-switched but non-injected, frequency doubled Nd:YAG laser emitting 15 ns pulses in a F/50 focusing geometry ; and (ii) an optical parametric oscillator (OPO) with a pulse duration of 3 ns at 532 nm, using a RSG-19 experimental setup in a F/5 focusing geometry with a 1.5 mrad collecting aperture.

## RESULTS AND DISCUSSION

### Relation between nanotube structure and optical limiting performances

The relation between nanotube structure and optical limiting properties was investigated by several groups so far, with contradictory results [6-10]. However, relevant conclusions can only be achieved on well-characterized samples [3,5]. We first studied the effect of the length of the nanotubes, using commercial sources of MWNT (from Nanolab, Inc), as well as SWNT, shortened by an oxidization treatment. We found that the length of the nanotubes has no effect on the optical limiting performances [10]. By contrast, we found that the diameter of the nanotubes plays an important role. In figure 1, we compare the limiting properties of aqueous suspensions of individual SWNT (diameter $\phi$ in the range 1.3-1.4 nm), bundled SWNT ($\phi$ in the range 10-15 nm), and MWNT ($\phi$ in the range 20-50 nm), in a F/50 focusing geometry. Both the limiting threshold and the maximal transmitted energy significantly decrease when the diameter increases.

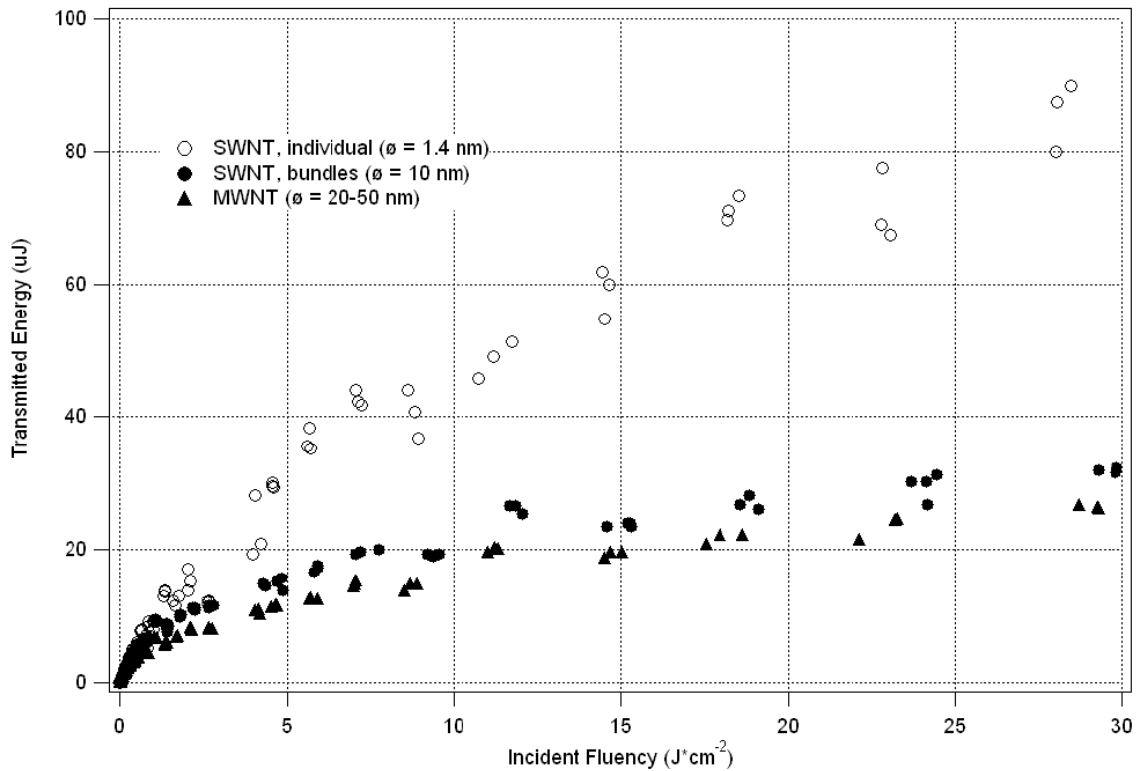

**Figure 1** : Optical limiting properties of various nanotube systems, with different diameters $\phi$ (532 nm, 15 ns, F/50).

In order to interpret these results, we first recall that the main mechanism responsible of optical limiting in suspensions of nanotubes is a strong nonlinear scattering due to heating of the nanotubes and the concomitant growth of solvent and/or carbon vapor bubbles [1]. Such a process can be decomposed in two steps : nucleation of scattering centers, and growth. From the Laplace law, one knows that the energy necessary for the nucleation of a bubble is inversely proportional to its diameter, which is expected to be related to the size of the heating center, *i.e.* the nanotube. Therefore, the nucleation of a bubble from large nanotubes will be energy-costless and will occur at lower laser fluences. Right after their nucleation, bubbles are too small to generate effective scattering. Growth of the bubbles occurs due to heating, but also to surpression (see evidence of cavitation effects in reference [1,11]) and nucleation with other bubbles [12]. The growth step, up to a size effective for scattering, will obviously be longer for smaller bubbles. Delays due to both nucleation and growth explain well the diameter-dependence of the limiting properties of suspended nanotubes. Finally, one can not rule out possible changes in the optical and/or thermal properties for the different nanotube systems. Such changes may modulate the kinetics of heat transfert from nanotubes to solvent and possibly that of nanotube sublimation. However, we believe that our experimental results clearly evidence the major role of the nanotube diameter in their optical performances.

## **Optical limiting from mixtures nanotubes / two-photon absorbers**

None system, taken individually, is able to fulfill the specifications of broadband optical limiting. Therefore, in order to extend the spectral and temporal ranges of effective limiting, different systems must be coupled. Such attempts were performed with combinations of non-linear optical materials in cascade geometries : multi-plate or tandem cells [13], or use of two intermediate focal planes of a sighting system [14]. We recently proposed and validated the concept of the combination of nanotubes and two-photon absorbers (TPA) in a same cell for effective broadband optical limiting [15,16]. Non linear scattering from nanotubes is powerful in the ns range and above. Non linear absorption from TPA is especially effective for ps pulses. The limiting performances of a mixture of SWNT and stilbene-3 (a commercial TPA) in aqueous supensions is significantly improved with respect to each single component alone in the ns range [15,16]. However, water is one of the poorest solvent for nucleation and growth of solvent vapor bubbles. On the other hand, raw nanotubes are not soluble in any solvent. Three main roads for preparing stable dispersions have been developed so far. The use of surfactants is a popular way to prepare aqueous suspensions. Grafting of organic functions on the tips or sidewalls of the nanotubes is another effective way to get stable dispersions [17]. Finally, it was demonstrated recently that *true* solutions of *charged* nanotubes can be prepared by protonation in superacids [18] or reduction (electron transfer) by alkali-atoms in DMSO and other polar solvents [19]. However, the nanotubes are likely individual (or assembled in small bundles) in these solutions. Moreover, the solvents which allow the preparation of solutions are still not very good solvents for the growth of solvent bubbles. Therefore, we chose the grafting way to make nanotubes soluble in chloroform. Grafting of octadecylamine chains on SWNT was achieved by C. Ménard, E. Doris and C. Mioskowski at CEA Saclay, France. The solubility of such functionnalized SWNT (SWNT-f) in chloroform is large enough to prepare effective samples for optical limiting experiments.

Figure 2 compares optical limiting results, in a RSG-19 geometry, for a mixture SWNT / FMP and each of the components alone. Its solubility in chloroform is high, we fixed 300 g.l$^{-1}$ for both the TPA alone and the mixture. For such large concentrations, the limiting properties of the TPA are still very high for 3 ns pulses, even better than those of the grafted-nanotubes at large fluences. The performances of the mixture are better than those of the two components alone, both at the limiting threshold and for the maximal transmitted energy.

# CONCLUSION

In this paper, we explored two possible routes for improving optical limiting performances of nanotube-based systems. We demonstrated that the nanotube diameter is a key factor : the larger the diameter and the better the limiting performances. Large bundles or large MWNT will therefore be good candidates for optical limiting, providing that stable suspensions can be prepared and keeping the sizes sufficiently small to prevent scattering at low fluences. On the other hand, we proposed and validated the concept of mixing nanotubes with two photon absorbers in the same cell, for high-performance, broadband limiting. Both the threshold and the maximal transmittance are optimized for ns pulses. Furthermore, good results are expected from ps to µs ranges, which makes composite nanotube-based systems ideal candidates for operational optical limiting systems.

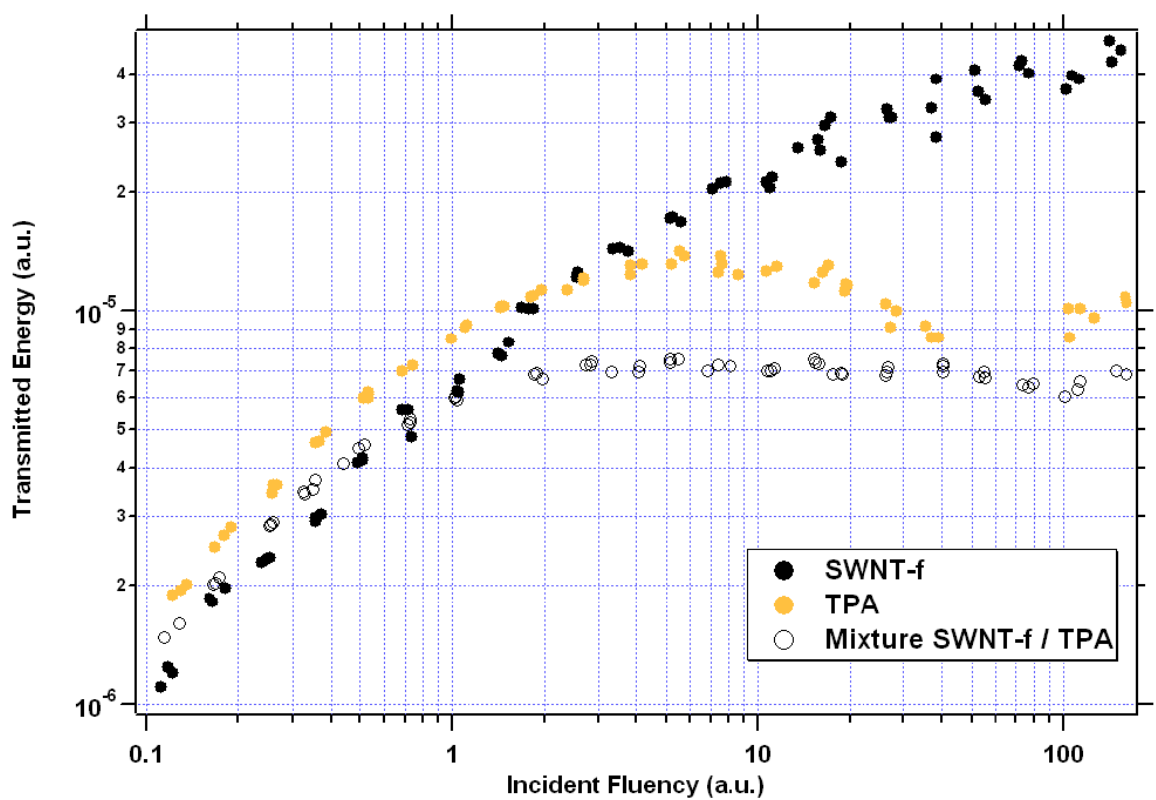

**Figure 2** : Optical limiting properties of a mixture of functionnalized SWNT (SWNT-f) and two-photon absorbers (TPA), for 3 ns pulses at 532 nm (RSG-19).

# ACKNOWLEDGEMENTS

We thank C. Ménard, E. Doris, C. Mioskowski from CEA Saclay and C. Andraud from Ecole Normale Supérieure de Lyon for providing high-quality samples and for fruitful intercations. E.A. acknowledges financial support from DGA.